\documentclass[aps,prb,amsmath,preprint,amssymb,titlepage,floatfix]{revtex4-2}
\usepackage{graphicx}
\usepackage{bm}
\usepackage{amssymb}
\usepackage{color}
\usepackage{amsmath}
\usepackage{multibib}

\usepackage[hypertexnames=false,colorlinks=true,linkcolor=blue,anchorcolor=black,citecolor=blue,filecolor=black,menucolor=black,runcolor=black,urlcolor=blue]{hyperref}
\newcites{A}{Main}
\newcites{B}{Methods}

\newcommand{\ignore}[1]{}

\begin{document}

\title{Boundary singular scattering enables coherent perfect absorption and lasing without resonators}

\author{Letian Yu}
\thanks{These authors contributed equally}
\affiliation{Division of Physics and Applied Physics, School of Physical and Mathematical Sciences, Nanyang Technological University, Singapore 637371, Singapore}

\author{Xu Zheng}
\thanks{These authors contributed equally}
\affiliation{Quantum Science Center of Guangdong-Hong Kong-Macao Greater Bay Area (Guangdong), Shenzhen 518045, China}
\affiliation{Division of Physics and Applied Physics, School of Physical and Mathematical Sciences, Nanyang Technological University, Singapore 637371, Singapore}

\author{Cesare Soci}
\email{csoci@ntu.edu.sg}
\affiliation{Division of Physics and Applied Physics, School of Physical and Mathematical Sciences, Nanyang Technological University, Singapore 637371, Singapore}
\affiliation{Centre for Disruptive Photonic Technologies, Nanyang Technological University, Singapore 637371, Singapore}

\author{Y. D. Chong}
\email{yidong@ntu.edu.sg}
\affiliation{Division of Physics and Applied Physics, School of Physical and Mathematical Sciences, Nanyang Technological University, Singapore 637371, Singapore}
\affiliation{Centre for Disruptive Photonic Technologies, Nanyang Technological University, Singapore 637371, Singapore}

\author{Baile Zhang}
\email{blzhang@ntu.edu.sg}
\affiliation{Division of Physics and Applied Physics, School of Physical and Mathematical Sciences, Nanyang Technological University, Singapore 637371, Singapore}
\affiliation{Centre for Disruptive Photonic Technologies, Nanyang Technological University, Singapore 637371, Singapore}

\maketitle

\clearpage
\begin{center}
{\bfseries Abstract}
\end{center}

\vspace{0.5em}

\noindent
Scattering singularities are features conventionally associated with resonant scattering bodies, which correspond to the zeros and poles of a scattering matrix, and manifest as coherent perfect absorption and lasing respectively.  Here, we show that they can emerge solely from an interface, eliminating the need for any resonant body. We introduce boundary singular scattering, whereby an interface between two non-Hermitian media supports coherent perfect absorption, lasing, or both. Despite these singular responses, the two adjoining media are in parity-time (PT)-unbroken phases, with entirely real bulk spectra. This phenomenon originates from a singular boundary mode-matching condition unique to non-Hermitian systems, not from any underlying resonant modes.  We demonstrate it experimentally in a programmable synthetic photonic lattice, and show coherent switching between perfect absorption and lasing by controlling only the relative phase of the incident beams. The same mechanism further generalizes from mirror-symmetric interfaces to asymmetric interfaces with dissimilar bulk dispersions. Our work establishes boundaries as a previously unexplored origin of scattering singularities, opening a resonator-free route to coherent perfect absorption, lasing, and other singular wave phenomena.
\clearpage

\section{Introduction}

The scattering of waves by obstacles is a topic of central interest in physics, with applications spanning fundamental physics \cite{ishimaruElectromagneticWavePropagation2017} and modern photonics \cite{chenAntireflectionCoatingUsing2010, moleskyInverseDesignNanophotonics2018, basarWirelessCommunicationsReconfigurable2019, kubbyWavefrontShapingBiomedical2019}. Among the most remarkable examples are scattering singularities, at which the scattering matrix becomes singular and the outgoing wave amplitudes either diverge or vanish relative to the incoming waves. In optics, these two limits correspond to lasing and coherent perfect absorption (CPA), the latter also known as ``anti-lasing'' \cite{cpa2010, cpaexpt2011, zhang2012cpa, baranov2017}. Under appropriate conditions, the two singularities can even coexist in a single system, forming a ``CPA-laser'' that can be switched between perfect absorption and lasing by varying the coherent excitation conditions \cite{longhiPTSymmetricLaser2010, chongSymmetryBreakingLaserAbsorber2011, mostafazadehSelfdualSpectralSingularities2012, baiSimultaneousRealizationCoherent2016, yangExperimentalObservationCoherentperfectabsorber2024a}. These singular wave responses arise from the combined action of optical gain or loss and coherent wave interference, making the realization of the required interference a central design question.

\begin{figure*}
  \centering
  \includegraphics[width=\textwidth]{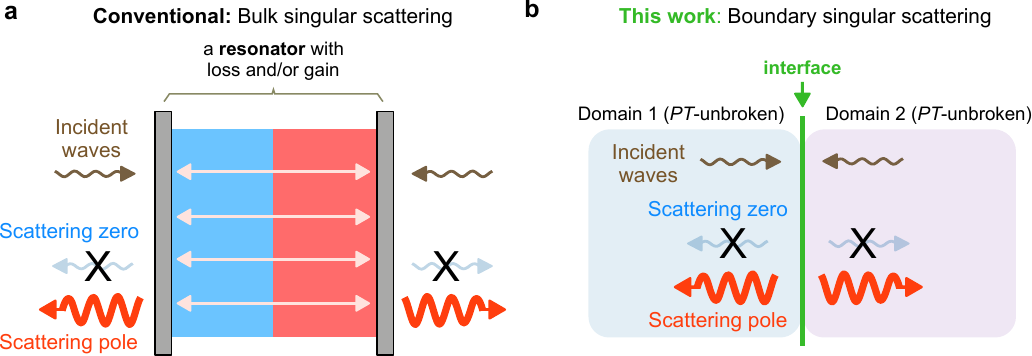}
  \caption{Schemes for realizing scattering singularities. \textbf{a}, Conventional resonator-based realization of scattering singularities. A two-port resonator containing gain and/or loss supports scattering zeros (coherent perfect absorption) and/or poles (lasing) at resonance. Waves propagating inside the resonator experience amplification or attenuation. \textbf{b}, Boundary singular scattering at the interface between two PT-symmetric media. The interface itself, rather than a separate resonant scattering body, supports scattering zeros and poles. Both adjoining media remain in the PT-unbroken phase, so propagating waves experience no net amplification or attenuation in the bulk.}
  \label{fig0}
\end{figure*}

Conventionally, scattering singularities arise from resonant wave confinement within a finite scattering body (Fig.~\ref{fig0}\textbf{a}). Lasing requires optical feedback provided by a resonator, while CPA, as its time-reversed counterpart, has likewise been conventionally realized through resonant scattering structures \cite{slobodkinMassivelyDegenerateCoherent2022, wangCoherentPerfectAbsorption2021, hornerCoherentPerfectAbsorption2024}. Coherent perfect absorber-lasers (CPALs) \cite{cpaexpt2011, mostafazadehSelfdualSpectralSingularities2012, baiSimultaneousRealizationCoherent2016, wongLasingAntilasingSingle2016a, yangExperimentalObservationCoherentperfectabsorber2024a, galiffiOpticalCoherentPerfect2026} impose even more stringent requirements by combining carefully engineered gain and loss within resonant structures \cite{geConservationRelationsAnisotropic2012, achilleosScatteringFinitePeriodic2017, leeWavePropagationBidirectional2023}, typically arranged with parity-time (PT) symmetry \cite{benderRealSpectraNonHermitian1998, benderPTsymmetricQuantumMechanics1999, benderComplexExtensionQuantum2002} and operating in the spontaneously PT-broken phase \cite{longhiPTSymmetricLaser2010, chongSymmetryBreakingLaserAbsorber2011, longhiPTsymmetricMicroringLaserabsorber2014, wongLasingAntilasingSingle2016a}. Resonant scattering structures have therefore remained the prevailing platform for realizing optical scattering singularities. This raises a fundamental question: can scattering singularities, including CPAL points, emerge without any resonant scattering body?

Here, we answer in the affirmative: scattering singularities can be realised through a fundamentally different route, bypassing the need for resonant scattering bodies. Specifically, an interface between two non-Hermitian media can sustain CPAL, while both adjoining media remain in the exact PT-unbroken phase with entirely real bulk spectra (Fig.~\ref{fig0}\textbf{b}). Although wave interference remains essential, it is established here through a singular boundary mode-matching condition unique to non-Hermitian systems rather than through resonant wave confinement. We refer to this phenomenon as boundary singular scattering. We experimentally realize this in a programmable synthetic photonic lattice, in which a single interface switches between CPA and lasing simply by varying the relative phase of the coherent excitations. Our work identifies boundaries as a previously unexplored origin of scattering singularities, establishing that the singular interference underlying CPA and lasing need not arise from resonant wave confinement. This opens a resonator-free route to coherent perfect absorption and lasing, and expands the design principles for singular wave manipulation in non-Hermitian systems.

\section{Results}
\subsection{Boundary scattering in a synthetic Floquet lattice}
\begin{figure*}
  \centering
  \includegraphics[width=\textwidth]{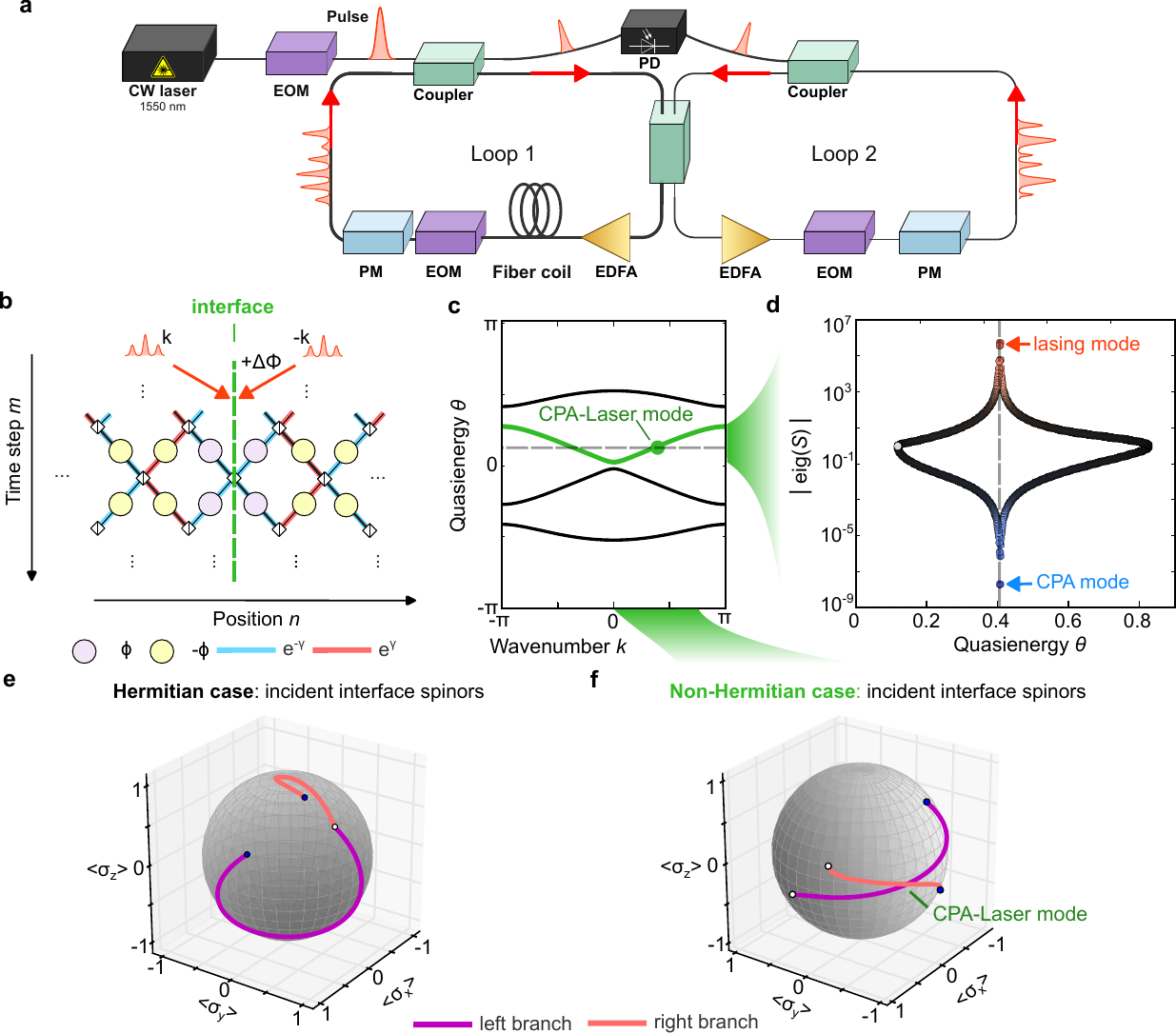}
  \caption{\textbf{Scattering from an interface in a non-Hermitian lattice.} \textbf{a}, Schematic of the setup comprising two coupled optical fiber loops. The system includes electro-optical modulators (EOM), erbium-doped fiber amplifiers (EDFA), phase modulators (PM) and photodetector (PD). \textbf{b} Mapping of the physical fiber loops onto a $1+1$D synthetic lattice, where time step $m$ and position $n$ govern the pulse evolution. The colored circles and shades indicate the applied phase factors ($e^{i\phi}, e^{-i\phi}$) and gain/loss factors ($e^{\gamma}, e^{-\gamma}$). The orange arrows denote the probe beams, with relative phase difference $\Delta\Phi$. \textbf{c}, Calculated Floquet band structure for the unit cell in the left domain. The green dot marks the CPAL mode. \textbf{d}, Absolute values of the scattering matrix eigenvalues for the green band in \textbf{c}. At the CPAL point, one eigenvalue diverges while the other vanishes.  \textbf{e}, Bloch sphere projected trajectories of the incident wavevectors from two sides of the interface when $\gamma = 0$ and $\phi = \pi/8$, as the corresponding momentum is swept from $k=0$ (marked by hollow circle) to $k=\frac{3}{4}\pi (-\frac{3}{4}\pi)$ (marked by filled circle) for the mode in the left (right) domain, respectively. \textbf{f}, the same as \textbf{e} but under non-Hermitian settings, where $\gamma = 0.382$ and $\phi = \pi/8$.}
  \label{fig1}
\end{figure*}

To test boundary singular scattering experimentally, we consider the platform of photonic mesh lattice (shown in Fig.~\ref{fig1}\textbf{a}), which describes a modified one-dimensional discrete-time light walk \cite{schreiberPhotonsWalkingLine2010, Regensburger2012, Miri2012a, weidemannTopologicalFunnelingLight2020, feisSpacetimetopologicalEventsPhotonic2025} based on coupled optical fiber loops. As the two loops have different lengths, pulses circulating in the system undergo repeated splitting and interference. This dynamics maps onto a $1+1$D Floquet lattice (Fig.~\ref{fig1}\textbf{b}) with the discrete evolution equations \cite{schreiberPhotonsWalkingLine2010}:
\begin{align}
  u_n^{m+1}&=\left(\cos\beta u_{n+1}^{m}+i\sin\beta v_{n+1}^{m}\right)e^{\gamma_n^{u}+i\phi_n^{u}}, \label{eq:evolution_equations0}
 \\
  v_n^{m+1}&= \left(i\sin\beta u_{n-1}^{m}+\cos\beta v_{n-1}^{m}\right)e^{\gamma_n^{v}+i\phi_n^{v}},
  \label{eq:evolution_equations}
\end{align}
where $v_n^m$ and $u_n^m$ are the pulse amplitudes in loop 1 (the right-moving branch in Fig.~\ref{fig1}\textbf{b}) and loop 2 (the left-moving branch) respectively, at lattice position $n$ and time step $m$.  The coupling coefficient $\beta=\pi/4$ parameterizes the 50:50 beamsplitter joining the loops. The imaginary potential is set by the gain/loss parameter $\gamma_n^{u/v}$ via the electro-optical modulator (EOM), and the real potential is set by $\phi^{u/v}_n$ via the phase modulator (PM)  \cite{Regensburger2012}. By driving these modulators with appropriate radio frequency (RF) signals, we can realize different complex potentials and boundary configurations and monitor the beam dynamics in both loops with photodetectors (PDs).

We now use this setup to configure PT symmetric bulk lattices.  First, consider the left domain in Fig.~\ref{fig1}\textbf{b}. Its unit cell spans four lattice sites, with a spatially antisymmetric imaginary potential produced by the gain/loss pattern $[\gamma, -\gamma, \gamma, -\gamma]$ and a symmetric real potential corresponding to the phase pattern $[\phi,-\phi,-\phi,\phi]$.  For each $\phi$, the bulk band structure is PT-unbroken (with all quasienergy bands entirely real) for $|\gamma| < \gamma_c(\phi)$, where $\gamma_c(\phi)$ is a critical gain/loss parameter; for $|\gamma| > \gamma_c(\phi)$, it enters a PT-broken phase with complex quasienergies (see Supplementary Information).  Next, we define the right domain by flipping the gain/loss pattern while keeping the phase profile unchanged, resulting in a mirror image of the left domain.  Each bulk domain is PT symmetric, even though the combined structure is not.

Suppose each bulk domain lies in the PT-unbroken phase, with a real quasienergy band structure. For the incident plane waves, we use a Floquet-Bloch ansatz to write the propagating modes in the left and right domains as
\begin{align}
\psi_l^m &= A_l\chi_l^+ e^{-\frac{i\theta m}{2}+\frac{ik(n-1)}{4}+\frac{im\pi}{2}} + R_l\chi_l^- e^{-\frac{i\theta m}{2}-\frac{ik(n-1)}{4}+\frac{im\pi}{2}}, \\
\psi_r^m &= A_r\chi_r^- e^{-\frac{i\theta m}{2}-\frac{ik(n-1)}{4}+\frac{im\pi}{2}} + R_r\chi_r^+ e^{-\frac{i\theta m}{2}+\frac{ik(n-1)}{4}+\frac{im\pi}{2}},
\end{align}
where $A_{l,r}$, $R_{l,r}$ are the incident and scattered amplitudes respectively. We determine $\chi_{l,r}^\pm$ by solving the Floquet-Bloch equations in each domain, with the $\pm$ superscripts denoting right- and left-propagating modes.  By wave-matching at the interface, we obtain the scattering relation (see Methods for details):
\begin{equation}
S_1 \begin{bmatrix} A_l e^{-ik} \\ A_r \end{bmatrix} = S_2 \begin{bmatrix} R_l e^{ik} \\ R_r \end{bmatrix}, \quad S = S_2^{-1} S_1.
\end{equation}
The eigenvalues of $S$ determine the CPAL condition: one eigenvalue vanishing implies perfect absorption, while the other simultaneously diverging implies lasing. Due the symmetries of $S$, each eigenvector has equal intensities in the two incidence channels. When the inter-channel phase difference $\Delta \Phi_{\textit{CPA}} = k_0$ for the CPA eigenvector, where $k_0$ corresponds to the momentum satisfying the CPAL condition, the corresponding lasing vector requires a phase difference of $\Delta \Phi_L = k_0 + \pi$ (see Methods).

For the exemplary phase parameter $\phi=\pi/8$, we find analytically that there is a specific $\gamma=0.382$ for which the bulk lattices are PT-unbroken (Fig.~\ref{fig1}\textbf{c}) and the CPAL condition is satisfied at a real quasienergy $\theta$ (Fig.~\ref{fig1}\textbf{d}).  More generally, we can locate CPAL solutions everywhere along a curve in the $(\gamma, \phi)$ parameter space, with $\theta$ also varying along the curve (see Supplemental information).  If we detune $\gamma$ and/or $\phi$ away from the curve, the CPA and lasing solutions individually persist, but move to separate values of $\theta$.

The emergence of scattering zeros and poles can be understood from the mode-matching of waves at the boundary. For a single boundary between Hermitian media, real-frequency scattering zeros or poles are prohibited by flux conservation and the associated propagation-direction constraints (see Supplemental Information). In the present Floquet lattice, this constraint can be visualized more directly by examining the wave amplitudes immediately before scattering at the interface. Specifically, we construct the two incident interface spinors from the amplitudes arriving at the coupler and project them onto the Bloch sphere (see Methods for more details). As shown in Fig.~\ref{fig1}\textbf{e}, under Hermitian settings when $\gamma=0$, the two counter-propagating incident spinors remain separated as the Bloch momentum is varied, except at the special point $k=0$. The situation changes once non-Hermiticity is introduced. Gain and loss modify the relative amplitudes and phases of the incoming waves, allowing the interface spinors associated with opposite incidence directions to collapse onto the same Bloch-sphere point while still retaining their opposite flux directions. For the exemplary parameters mentioned above ($\phi=\pi/8$ and $\gamma=0.382$), the two trajectories of the incident interface spinors intersect as the Bloch momentum is swept over $k\in[0,\frac{3}{4}\pi]$, as shown in Fig.~\ref{fig1}\textbf{f}. At this intersection, the two incident spinors become linearly dependent precisely at the interface. This is the spinor-level signature of the singular mode-matching condition derived above, which is only possible here due to the presence of non-Hermiticity.


\subsection{Experimental observation of boundary CPAL}

\begin{figure*}
  \centering
  \includegraphics[width=0.8\textwidth]{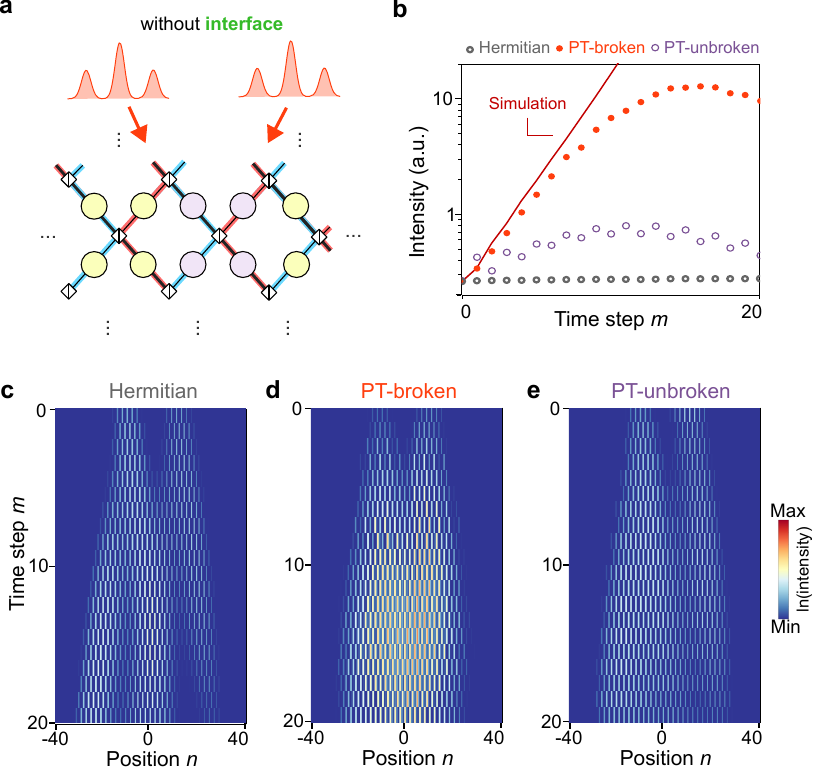}
  \caption{\textbf{Wave propagation in a uniform PT-symmetric lattice.} \textbf{a}, Schematic of the PT-symmetric Floquet lattice, initialized with two coherent Gaussian wavepackets.  \textbf{b},  Graph of total intensity versus time obtained experimentally for a Hermitian lattice ($\gamma=0$, gray circles), a PT-broken lattice ($\gamma = 0.382$ and $\phi = 0$, red circles), and a PT-unbroken lattice ($\gamma = 0.382$ and $\phi = \pi/8$, purple circles).  The intensities are conserved in the Hermitian lattice and approximately so in the PT-unbroken lattice, whereas in the PT-broken lattice it increases exponentially, consistent with the simulated curve (red line).  \textbf{c}--\textbf{e}, Experimentally-obtained beam dynamics data for the three synthetic lattices shown in \textbf{b}: Hermitian (\textbf{c}), PT-broken (\textbf{d}), and PT-unbroken (\textbf{e}). The color scale represents the natural logarithm of the intensity. }
  \label{fig2}
\end{figure*}

We now implement the boundary CPAL setup experimentally.  As an initial calibration step, we study the evolution of counterpropagating wavepackets in a uniform, boundary-free lattice (Fig.~\ref{fig2}).   In Fig.~\ref{fig2}\textbf{b}--\textbf{e}, we present the experimental data for three lattice configurations: a Hermitian lattice ($\gamma = \phi = 0$), a PT-broken lattice ($\gamma=0.382$ and $\phi = 0$), and a PT-unbroken lattice ($\gamma=0.382$ and $\phi=\pi/8$).   For these three cases, the total intensity respectively remains constant, grows exponentially, and remains stable, both before and after the wavepackets collide.  Notably, the exponential growth observed in the PT-broken case initially closely matches simulation results (Fig.~\ref{fig2}\textbf{b}), but eventually rolls over due to amplifier saturation and recovery \cite{Wimmer2015}.

The picture changes markedly when we introduce the boundary by flipping the gain-loss pattern over half the lattice (Figure~\ref{fig1}\textbf{b}). The two incident Gaussian wavepackets are now prepared with central wavevectors $\pm k$ (see Methods), and a relative phase $\Delta\Phi$ is added to the left-moving packet and controlled by the PMs in the preceding round trip.

\begin{figure*}
  \centering
  \includegraphics[width=\textwidth]{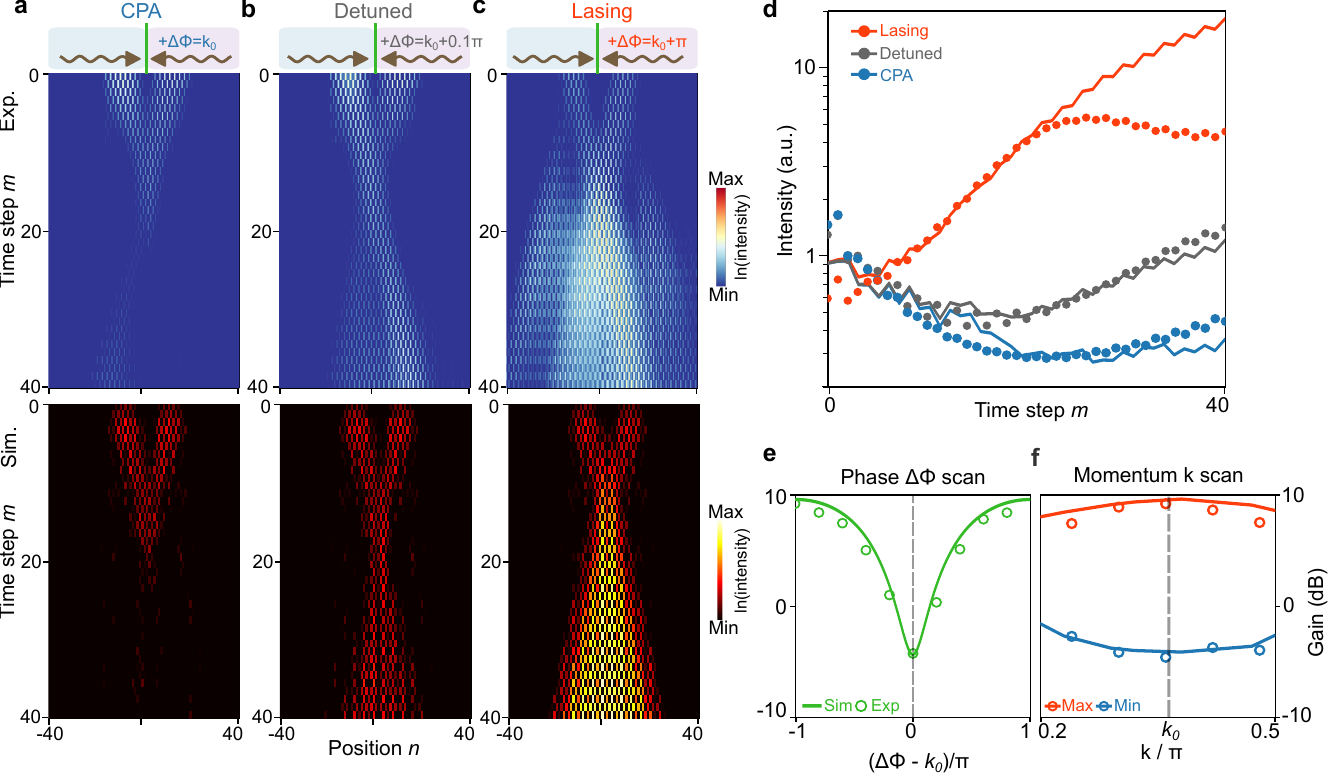}
  \caption{\textbf{Experimental observation of boundary CPAL.} \textbf{a-c}, Experimental (left) and simulated (right) beam dynamics for the configuration in Figure~\ref{fig1}b with relative phases for \textbf{a} --  $\Delta\Phi = k_0$ (CPA regime), \textbf{b} --  $\Delta\Phi = k_0 + 0.1\pi$ (detuned regime), and \textbf{c} --  $\Delta\Phi = k_0 + \pi$ (lasing regime). The color scale maps the natural logarithm of the intensity.
  \textbf{d}, Extracted intensity evolution versus time step $m$ for the corresponding phase differences in (a-c). The solid curves denote simulations, and the points denote experimental measurements. \textbf{e}, Total intensity ratio (in dB) evaluated over a 20-step window as a function of phase difference. The dip and peak correspond to the CPA and lasing limits, respectively. \textbf{f}, Maximum and minimum intensity ratios obtained for selected normalized wavevector $k/\pi$. The dashed vertical line indicates the primary operating wavevector $k_0$, which correspond to the CPAL frequency.}
  \label{fig3}
\end{figure*}

In Fig.~\ref{fig3}\textbf{a}--\textbf{c}, we show the results for fixed $k = k_0 \equiv 0.364\pi$ and different $\Delta \Phi$.  When $\Delta \Phi = k_0$, we observe that after the wavepackets meet from the boundary, the scattered intensity is strongly suppressed (Fig.~\ref{fig3}a), consistent with the preceding theoretical predictions.  A slight intensity revival is observed at much later times, which we attribute to the residual excitation of modes in other bands.  A small phase detuning away from the CPA condition ($\Delta\Phi=k_0+0.1\pi$) restores much of the scattered intensity (Fig.~\ref{fig3}\textbf{b}).  If we set the phase difference to $\Delta \Phi = k_0+\pi$, we instead observe strong amplification (Fig.~\ref{fig3}\textbf{c}).  These results confirm the existence of the boundary CPAL phenomenon in this lattice.  The measured evolution of the total intensity (summed over all lattice positions) with time is shown in Figure~\ref{fig3}\textbf{d}; the data agree closely with numerical predictions, aside from the gain saturation-induced rollover for the lasing solution.

We can also characterize the CPA and lasing solutions by varying the wavevector $k$.  In Fig.~\ref{fig3}\textbf{e},\textbf{f}, we show experimental data for how the output coefficient of the CPA laser, defined as the ratio of total scattered intensity (measured at $m=24$) to the incoming light intensity (measured at $m=0$), varies with either $\Delta \phi$ or $k$, with the other parameter fixed. Although the contrast is reduced by the finite width of the incident Gaussian packets, the measured CPAL position agrees well with the predictions from the numerical simulations.

\subsection{Boundary CPAL under asymmetric incidence}

\begin{figure*}
  \centering
  \includegraphics[width=0.9\textwidth]{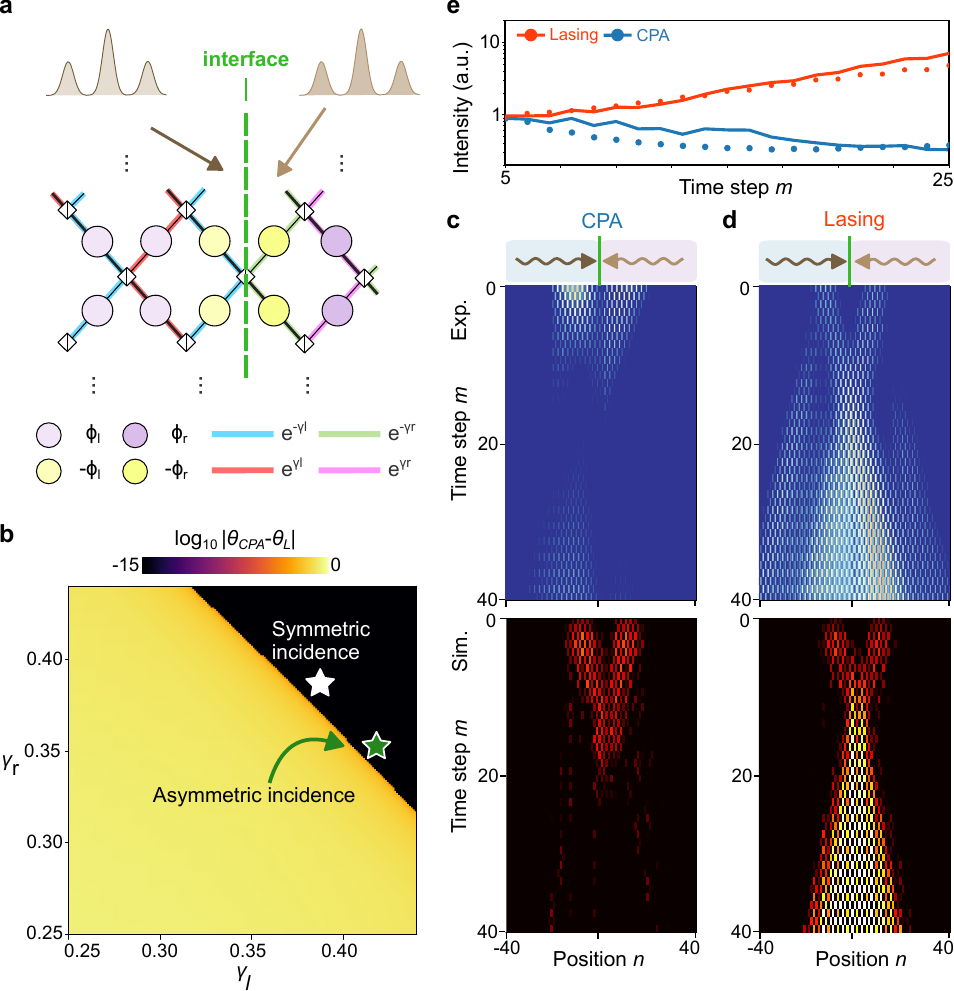}
  \caption{\textbf{Boundary CPAL under asymmetric incidence.} \textbf{a}, Schematic of two adjacent PT-symmetric domains coupled at an interface and implemented with coupled fiber loops. Different color shades indicate different EOM and PM modulation strengths. \textbf{b}, Phase diagram of the quasienergy separation between the CPA and lasing points in the parameter space spanned by the gain/loss amplitudes in the left and right domains ($\gamma_l$ and $\gamma_r$). White (green) stars indicate the representative parameter choices used here for symmetric (asymmetric) incidence in the experiment. \textbf{c,d}, Experimental (left) and simulated (right) beam dynamics under asymmetric illumination for two different phase differences. \textbf{e}, Extracted intensity evolution versus time step $m$ for the corresponding phase differences in \textbf{c} and \textbf{d}. The solid curves denote simulations, and the points denote experimental measurements.}
  \label{fig4}
\end{figure*}

In the configurations studied so far, the two domains are mirror images of each other, with identical bulk dispersions, and the symmetry of the scattering matrix constrains the CPA and lasing eigenvectors to have equal intensities in the two incidence channels. A natural question is whether boundary singular scattering relies on this symmetry. We therefore consider asymmetric configurations, in which the two sides of the interface are assigned different complex potentials $(\gamma_l, \phi_l)$ and $(\gamma_r, \phi_r)$ and hence dissimilar bulk dispersions, as shown in Fig.~\ref{fig4}\textbf{a}. Beyond probing the generality of the boundary mechanism, such configurations realize asymmetric interferometric light-light control, which previously required engineered asymmetric media \cite{nefedovTotalAbsorptionAsymmetric2013}, exceptional points \cite{zhaoMetawaveguideAsymmetricInterferometric2016a} or time-varying materials \cite{galiffiOpticalCoherentPerfect2026}, but is obtained here by simply reprogramming the lattice modulation.  We scan different combinations of $\gamma_l$ and $\gamma_r$ and, for each pair, numerically minimize the separation between the lasing and CPA quasienergies with respect to $\phi_l$ and $\phi_r$.

Figure~\ref{fig4}\textbf{b} reveals a broad region of parameter space in which CPAL is attainable. Since the dispersion relations in the two domains are no longer identical, the CPAL condition is asymmetric: the two incident wavepackets generally have different central wavevectors, amplitudes and relative phase, and singular scattering occurs when they are exactly matched at the interface (see Supplementary Information).

We experimentally test a representative configuration with $\gamma_{l}=0.43$ and $\gamma_{r}=0.345$. The measured intensity dynamics again show strong attenuation and amplification for the two corresponding input phases (Figure~\ref{fig4}\textbf{c},\textbf{d}), in good agreement with the simulations. These results show that the same boundary mechanism persists after the two domains are reconfigured and that CPAL behavior is not limited to the symmetric-incidence case.

\section{Conclusion}
We have demonstrated theoretically and experimentally that scattering singularities can emerge directly from an interface between two non-Hermitian media, not requiring any resonant scattering body. In contrast to the conventional picture, whereby CPA and lasing emerge from resonant wave confinement, boundary singular scattering arises from an boundary matching condition unique to non-Hermitian systems. Consequently, a single interface can support both CPA and lasing, even though both adjoining media remain in the exact PT-unbroken phase with entirely real bulk spectra.  We further showed that the same mechanism extends to non-mirror-symmetric configurations, with dissimilar bulk dispersion relations on either side of the interface. This implies that boundary singular scattering is a general interfacial phenomenon that does not rely on global symmetry or a particular lattice realization.

More fundamentally, our work identifies interfaces as an under-explored source of extreme scattering phenomena. This shifts the physical picture of scattering singularities from one centered on resonant scattering bodies to one in which they can be engineered directly through non-Hermitian boundaries. Beyond providing a resonator-free route to CPA and lasing, our results broaden the physical mechanisms available for engineering singular wave phenomena and establish interface engineering as a new strategy for singular wave manipulation in non-Hermitian systems.

\section{Methods}
\subsection{Experimental measurements}

In our experimental setup (Fig.~\ref{fig1}\textbf{a}), each loop comprises a spool of single-mode fiber with an average round-trip time of approximately 20 $\mu$s. The length difference between the two loops corresponds to approximately 300 ns.  Each discrete-time walk is initiated using a square pulse of 100 ns, cut from a 1550 nm continuous-wave (CW) laser source. The square pulse is injected into the long loop and thereafter undergoes repeated splitting and interference in the two loops, forming a pulse chain.

To prepare two counter-propagating beams, we retain only the pulses at the two ends of the pulse chain, suppressing all the others using intensity modulators. We then apply a diffusion protocol to shape these pulses into the desired Gaussian beams \cite{Miri2012a}. We monitor the pulse dynamics in both loops by outcoupling a small portion of the optical signal into photodetectors connected to an oscilloscope.

\subsection{Scattering matrix}

Here, we describe the derivation of the scattering matrix for a boundary between two bulk lattice domains (Fig.~\ref{fig1}\textbf{b}). First, we consider each individual domain: the modulation pattern has temporal period $\Delta m=2$ and spatial period $\Delta n=4$, so it is convenient to work with a two-step evolution.  The evolution equations \eqref{eq:evolution_equations} are thus expanded into
\begin{align}
  u_n^{m+2}&=\frac{1}{2}\left[\left(u_{n+2}^{m}+i v_{n+2}^{m}\right)e^{-2i\phi} + \left(-u_{n}^{m}+i v_{n}^{m}\right)e^{-2i\phi-2\gamma}\right], \label{eq:evo1} \\
  v_n^{m+2}&=\frac{1}{2}\left[\left(iu_{n}^{m}- v_{n}^{m}\right)e^{2i\phi+2\gamma} + \left(iu_{n-2}^{m}+ v_{n-2}^{m}\right)e^{2i\phi}\right], \\
  u_{n+2}^{m+2}&=\frac{1}{2}\left[\left(u_{n+4}^{m}+i v_{n+4}^{m}\right)e^{2i\phi} + \left(-u_{n+2}^{m}+i v_{n+2}^{m}\right)e^{2i\phi-2\gamma}\right], \\
  v_{n+2}^{m+2}&=\frac{1}{2}\left[\left(iu_{n+2}^{m}- v_{n+2}^{m}\right)e^{-2i\phi+2\gamma} + \left(iu_{n}^{m}+ v_{n}^{m}\right)e^{-2i\phi}\right]. \label{eq:evo4}
\end{align}
We seek Floquet-Bloch solutions of the form
\begin{align}
  \begin{pmatrix}
  u_n^m \\
  v_n^m \\
  u_{n+2}^m \\
  v_{n+2}^m
  \end{pmatrix}=\chi\, e^{-\frac{im\theta}{2}+\frac{ik(n-1)}{4}+\frac{im\pi}{2}},
  \;\; \chi \equiv \begin{pmatrix}
    U_0 \\
    V_0 \\
    U_1 \\
    V_1
  \end{pmatrix},
\end{align}
where $\theta$ is the quasienergy and $k$ is the wavevector. Substituting this into Eqs.~\eqref{eq:evo1}--\eqref{eq:evo4} yields an eigenvalue problem of the form
\begin{align}
  W(\gamma,\phi,k) \, \chi = e^{-i\theta} \,\chi,
	\label{eq:eigenvalue dispersion}
\end{align}
which leads to the dispersion relation
\begin{align}
  \cos{\theta}=\frac{1}{2}\cos{2\phi}\,\cosh{2\gamma}
  \pm\frac{1}{4} \sqrt{5+2\cos{k}-3\cos{4\phi}+(\cos{4\phi}-1)\cosh{4\gamma}},
    \label{eq:dispersion}
\end{align}
Here the $\pm$ sign gives two branches of the dispersion, each of which can have $\pm\theta$ solutions.  This gives four bands in total, which are plotted in Fig.~\ref{fig1}\textbf{c}.  Next, we consider two uniform domains, denoted left ($l$) and right ($r$).  In each domain, for a given propagating band let $\chi_{l,r}^\pm$ denote the eigenvectors with wavevector $\pm k$ at quasienergy $\theta$, and construct a superposition of propagating waves:
\begin{align}
\psi_l^m &= A_l\chi_l^+ e^{-\frac{i\theta m}{2}+\frac{ik(n-1)}{4}+\frac{im\pi}{2}} + R_l\chi_l^- e^{-\frac{i\theta m}{2}-\frac{ik(n-1)}{4}+\frac{im\pi}{2}}, \\
\psi_r^m &= A_r\chi_r^- e^{-\frac{i\theta m}{2}-\frac{ik(n-1)}{4}+\frac{im\pi}{2}} + R_r\chi_r^+ e^{-\frac{i\theta m}{2}+\frac{ik(n-1)}{4}+\frac{im\pi}{2}}.
\end{align}
We now place the interface at $n=0$. (Since the interface involves the intermediate site at time step $m+1$, while the Floquet-Bloch ansatz is formulated for the two-step evolution ending at $m+2$, there is a choice of unit cell convention. We may either introduce the shifted basis $((u_{n+1},v_{n+1},u_{n+3},v_{n+3})e^{ikn/4})$ or use the above shifted phase factor.) Matching the waves across a two-step evolution gives
\begin{align}
  u_{-1}^{m+2} &= \frac{1}{2}\left[\left(u_{1}^{m}+i v_{1}^{m}\right) + \left(-u_{-1}^{m}+i v_{-1}^{m}\right)\right]e^{2i\phi-2\gamma}, \\
  v_{1}^{m+2} &= \frac{1}{2}\left[\left(iu_{1}^{m}- v_{1}^{m}\right) + \left(iu_{-1}^{m}+ v_{-1}^{m}\right)\right]e^{2i\phi-2\gamma}.
\end{align}
Substituting the propagating modes into these conditions yields the scattering relation
\begin{align}
S_1 \begin{bmatrix} A_le^{-ik} \\ A_r \end{bmatrix} = S_2 \begin{bmatrix} R_le^{ik} \\ R_r \end{bmatrix}, \quad S = S_2^{-1} S_1,
\end{align}
where
\begin{align}
  S_1 &= \begin{pmatrix} 
    -e^{-i\theta}U_{1,l}^+ + \frac{1}{2}\left(U_{1,l}^+-iV_{1,l}^+\right)e^{2i\phi-2\gamma} & -\frac{1}{2}\left(U_{0,r}^- + iV_{0,r}^-\right)e^{2i\phi-2\gamma} \\
    -\frac{1}{2}\left(iU_{1,l}^+ + V_{1,l}^+\right)e^{2i\phi-2\gamma} & -e^{-i\theta}V_{0,r}^- + \frac{1}{2}\left(-iU_{0,r}^-+V_{0,r}^-\right)e^{2i\phi-2\gamma}
  \end{pmatrix} \\
  S_2 &= \begin{pmatrix} 
    e^{-i\theta}U_{1,l}^- - \frac{1}{2}\left(U_{1,l}^--iV_{1,l}^-\right)e^{2i\phi-2\gamma} & \frac{1}{2}\left(U_{0,r}^+ + iV_{0,r}^+\right)e^{2i\phi-2\gamma} \\
    \frac{1}{2}\left(iU_{1,l}^- + V_{1,l}^-\right)e^{2i\phi-2\gamma} & e^{-i\theta}V_{0,r}^+ - \frac{1}{2}\left(-iU_{0,r}^++V_{0,r}^+\right)e^{2i\phi-2\gamma}
  \end{pmatrix}
\end{align}
The CPAL effect occurs when $S_1$ and $S_2$ become singular simultaneously, so that one scattering eigenvalue vanishes while the other diverges. Requiring $\det S_1=\det S_2=0$ gives
\begin{align}
    \tanh(2\gamma)&=\tan{\frac{k}{2}},\\
    \tan{(2\phi)}&=\pm\frac{\sin{k}}{\sqrt{\cos^2{k}+4\cos{k}-1}},\\
     \tan{\theta}&=\tan{(2\phi)}\tan^{2}{\frac{k}{2}}.
\end{align}
Thus, the CPAL condition can be expressed parametrically in terms of the momentum $k$. Once the $k$ is chosen within the allowed range, the equations above fix the corresponding values of $\gamma, \phi, \theta$ for simultaneous coherent perfect absorption and lasing.

\subsection{Eigenvector spinor structure at CPAL point}

In the main text, we have shown that the emergence of scattering zeros and poles can be understood from the mode-matching of waves at the boundary. Here, we provide the proof in the synthetic Floquet lattice: at the CPAL point, the two incident spinors become linearly dependent at the interface. We start by substituting the bulk dispersion relation Eq.~\eqref{eq:eigenvalue dispersion} into the entries of $S_1$. For the incident mode from the left domain, the eigenvalue equation gives
\begin{equation}
-e^{-i\theta}U_{1,l}^+ = \frac{1}{2}\left[\left(U_{0,l}^++i V_{0,l}^+\right)e^{2i\phi+ik} + \left(-U_{1,l}^++i V_{1,l}^+\right)e^{2i\phi-2\gamma}\right]
\end{equation}
Similarly, for the incident mode from the right domain, where both $\gamma$ and $k$ change sign, one obtains
\begin{equation}
-e^{-i\theta}V_{0,r}^-=\frac{1}{2}\left[\left(iU_{0,r}^-- V_{0,r}^-\right)e^{2i\phi-2\gamma} + \left(iU_{1,r}^-+ V_{1,r}^-\right)e^{2i\phi+ik}\right]
\end{equation}
Using these relations, the matrix $S_1$ can be rewritten as
\begin{equation}
S_1
=
\frac{1}{2}e^{2i\phi}
\begin{pmatrix}
e^{ik}\left(U_{0,l}^+ + iV_{0,l}^+\right)
&
-e^{-2\gamma}\left(U_{0,r}^-+iV_{0,r}^-\right)
\\[4pt]
-e^{-2\gamma}\left(iU_{1,l}^+ + V_{1,l}^+\right)
&
e^{ik}\left(iU_{1,r}^-+V_{1,r}^-\right)
\end{pmatrix}.
\end{equation}
where the scattering zero corresponds to the quantity
\begin{align}
\det S_1
&=
\frac{1}{4}e^{4i\phi}
\left[
\left(U_{0,l}^+ + iV_{0,l}^+\right)
\left(iU_{1,r}^-+V_{1,r}^-\right)
-
e^{-4\gamma-2ik}
\left(U_{0,r}^-+iV_{0,r}^-\right)
\left(iU_{1,l}^+ + V_{1,l}^+\right)
\right].
\end{align}
being 0. Equivalently, the zero condition can be written as a matching condition between the two incident amplitude ratios at the interface:
\begin{equation}
\frac{
(U_{0,l}^+ + iV_{0,l}^+)e^{\gamma+ik}
}{
(iU_{1,l}^+ + V_{1,l}^+)e^{-\gamma}
}
=
\frac{
(U_{0,r}^-+iV_{0,r}^-)e^{-\gamma}
}{
(iU_{1,r}^-+V_{1,r}^-)e^{\gamma+ik}
}.
\label{eq:interface_spinor_matching}
\end{equation}
The numerator and denominator in Eq.~\eqref{eq:interface_spinor_matching} are precisely the two components of the incident waves after they are propagated to the interface. Therefore, the scattering zero is equivalent to the condition that the two incident spinors become linearly dependent at $n=0$. Explicitly, the condition can be expressed as the linear dependence between these two vectors
\begin{equation}
\begin{pmatrix}
e^{ik}u_{-4}\\
v_0
\end{pmatrix},
\qquad
\begin{pmatrix}
e^{-ik}u_0\\
v_4
\end{pmatrix}
\end{equation}
This gives a direct spinor-level interpretation of the CPAL condition: the singularity of the scattering matrix occurs when the two incoming channel modes collapse onto the same interface spinor direction.

This criterion can be checked directly after substituting in the corresponding bulk eigenvector components. In Fig.~\ref{fig1}\textbf{f}, we project the two incident interface spinors onto the Bloch sphere. Their intersection coincides with the CPAL point obtained independently from the singularity of the scattering matrix, confirming that the CPAL response originates from the linear dependence of the incident channel spinors at the interface.

\subsection{Excitation conditions for CPA and lasing modes}

The excitation condition for the CPA channel follows from the null eigenvector of $S_1$
\begin{equation}
S_1 \begin{bmatrix} A_le^{-ik} \\ A_r \end{bmatrix} = 0
\label{eq:cpa_condition_S1}
\end{equation}
At the CPAL point, $\det S_1=0$, so Eq.~\eqref{eq:cpa_condition_S1} has a nontrivial solution. Using the parity relation between right and left domains,
\begin{equation}
\chi_r^{-}=\mathcal{P}\chi_l^+,
\end{equation}
the diagonal terms, as well as the off-diagonal terms of $S_1$, become equal to each other in this case. The CPA condition is therefore
\begin{equation}
A_le^{-ik}/A_r = \pm1
\end{equation}
Thus the two incident waves must have equal amplitudes, while their relative phase is fixed to be either $k$ or $k+\pi$. Substituting the CPA condition back into the first row of $S_1$ gives
\begin{equation}
e^{-i\theta_{\pm}}U_{1,l}^+ = \frac{1\pm i}{2}\left(U_{1,l}^+-iV_{1,l}^+\right)e^{2i\phi-2\gamma}
\end{equation}
where $\theta_+-\theta_-=\frac{\pi}{2}$. Clearly, only one of the solution corresponds to a physical CPA state of the lattice. 

The lasing mode itself is a notrivial outgoing solution without incident field, which is generally independent of the excitation condition. The largest output intensity is obtained when the CPA channel is suppressed the most. In the present CPAL case, this corresponds to a state that has minimum overlap with the CPA mode. Hence, the excitation mode for the lasing condition also has equal amplitudes, with an additional $\pi$ phase shift from the CPA condition.

\section{Acknowledgements}

This work was supported by the Singapore National Research Foundation (NRF) under Competitive Research Program (CRP) NRF-CRP29-2022-0003 and the NRF Investigatorship NRF-NRFI08-2022-0001.

\bibliography{references}

\clearpage

\end{document}